\documentclass{elsart}

\usepackage{graphicx,amssymb,amsmath}
\usepackage{graphicx,color}
\begin{document}

\newcommand{\BQ}{\begin{equation}}
\newcommand{\EQ}{\end{equation}}
\newcommand{\BQA}{\begin{eqnarray}}
\newcommand{\EQA}{\end{eqnarray}}
\newcommand{\be}{\begin{eqnarray}}
\newcommand{\ee}{\end{eqnarray}}
\newcommand{\NN}{\nonumber \\}
\newcommand{\del}{\partial}
\def\simge{\mathrel{%
   \rlap{\raise 0.511ex \hbox{$>$}}{\lower 0.511ex \hbox{$\sim$}}}}
\def\simle{\mathrel{
   \rlap{\raise 0.511ex \hbox{$<$}}{\lower 0.511ex \hbox{$\sim$}}}}

\begin{frontmatter}

\begin{flushright}
UT-Komaba/09-2\\
KEK-TH-1304\\
\end{flushright}

\vspace{1.cm}

\title{Instabilities in non-expanding glasma}

\author{Hirotsugu Fujii}
\address{Institute of Physics, University of Tokyo, Komaba,
         Tokyo 153-8902, Japan,\\
         Yukawa Institute of Theoretical Physics, Kyoto 
         University, Kyoto 606-8502, Japan }
\author{Kazunori Itakura}
\address{Institute of Particle and  Nuclear Studies,
         High Energy Accelerator Research Organization (KEK),
         Tsukuba, Ibaraki 305-0801, Japan,\\
         Yukawa Institute of Theoretical Physics, Kyoto 
         University, Kyoto 606-8502, Japan}
\author{Aiichi Iwazaki}
\address{International Economics and Politics, 
         Nishogakusha University,\\ 
         Ohi, Kashiwa, Chiba 277-8585, Japan.}   

\date{May 27, 2009}

\begin{abstract}
A homogeneous color magnetic field is known to be unstable 
for the fluctuations perpendicular to the field in the color 
space (the Nielsen-Olesen instability). We argue that these 
unstable modes, exponentially growing, generate an 
{\it azimuthal} magnetic field 
with the original field being in the $z$-direction, 
which causes
the Nielsen-Olesen instability for another type of fluctuations. 
The growth rate of the latter unstable mode increases with 
the momentum $p_z$ and can become larger than 
the former's growth rate which decreases with increasing $p_z$.
These features 
may explain the interplay between the primary and secondary 
instabilities observed in the real-time simulation of a 
non-expanding glasma, i.e., stochastically generated 
anisotropic Yang-Mills fields without expansion.
\end{abstract}

\end{frontmatter}

\maketitle

\newpage
\section{Introduction}

Apparent success of hydrodynamic approaches in describing bulk 
properties of the data in relativistic heavy-ion collisions at 
BNL-RHIC \cite{HIC} has stimulated great interest in 
understanding the mechanism for (possible) short-time thermalization
of the system from the initial colliding stage. 
Preceding studies on
this subject \cite{review_Itakura} 
include the scenario based on the plasma instability 
 (in particular, the Weibel instability) 
caused by coupling between hard particles and soft fields 
\cite{Plasma_instability}, 
the `bottom-up' scenario based on the perturbative scatterings 
of hard particles \cite{bottom-up}, and so on. 
In addition to them, we have recently proposed a novel scenario 
based on the Nielsen-Olesen (N-O) instability \cite{nielsen} 
which is characteristic of the configuration of a uniform 
magnetic field in non-Abelian gauge theories 
\cite{Iwazaki2008,FujiiI2008}. This scenario applies the 
{\it earliest} time evolution of the system in the heavy ion 
collision when the separation between hard and soft degrees 
of freedom is not clear.

In the high-energy limit of collisions, the incident nuclei 
can be treated as two sheets of classical transverse gluonic 
fields in the Color Glass Condensate (CGC) framework \cite{CGC}. 
This is the QCD effective theory for dense small-$x$ degrees 
of freedom with treating the large-$x$ components as random 
colored sources moving on the light cone. With the initial 
condition given by CGC, it has been shown that {\it longitudinal} 
color electric and magnetic fields are produced just after two 
infinitesimally thin nuclei pass through each other \cite{Glasma_basic}. Since the correlation length of the fields on the transverse plane 
is typically $\sim 1/Q_s$ with $Q_s$ being the saturation scale of 
colliding nuclei, the emerging configuration after the 
collision is made of many longitudinal color flux tubes. 
The produced gluonic system evolving from this unique 
configuration towards locally thermalized state, providing 
an equilibrated plasma will be formed, is called {\it glasma} 
\cite{LappiM2006}.

The system of this classical field remains longitudinally
boost-invariant during time evolution even though it rapidly 
expands in the transverse plane and is stretched out in the 
beam direction. Especially, positive longitudinal pressure 
is never generated \cite{LappiM2006,RomatschkeV2005},
and therefore no thermalization is expected to be achieved.
In reality, however, boost-invariance of the system is 
naturally violated due to finite thickness of the colliding
nuclei and quantum fluctuations \cite{FukuGM2006} both of 
which are in general rapidity dependent. 
These rapidity-dependent effects may be treated as small fluctuations 
on top of the boost-invariant classical configurations.
The first investigation on the consequences of 
rapidity-dependent fluctuations was attempted in a numerical 
simulation of the classical SU(2) Yang-Mills field in an expanding 
geometry (i.e., in the proper-time and rapidity 
$\tau$-$\eta$ coodinates) \cite{RomatschkeV2005} and it was shown that 
the soft fluctuations indeed grow exponentially,
contributing positively to the longitudinal pressure, 
although the adopted initial fluctuations seemed somewhat 
too small. This simulation strongly suggests that the 
boost-invariant system is {\it unstable} with respect 
to the rapidity-dependent soft fluctuations.

Later, analytic studies of the instabilities in the glasma 
were presented in our works\cite{Iwazaki2008,FujiiI2008} within
simplified configurations for the background fields. We 
revealed that the characteristic features of the 
rapidity-dependent fluctuations found numerically in 
Ref.~\cite{RomatschkeV2005} are qualitatively reproduced 
as the instability of the homogeneous color magnetic field 
\`a la Nielsen-Olesen \cite{nielsen}. For example, the maximum 
longitudinal momentum of the unstable modes contributing to 
the pressure is shown to increase linearly with the proper-time 
$\tau$ in accord with the result in Ref.~\cite{RomatschkeV2005}.

Recently, another unstable behavior in the classical SU(2) 
gauge field simulation was reported in Ref.~\cite{BergesSS2007} 
(see also Ref.~\cite{BergesGSS2008} for SU(3) case). 
In this simulation, it was observed that 
(i) there are primary and secondary instabilities, the former 
of which is identified with the Weibel instability by the authors,
and that (ii) the secondary growth rate is larger than the 
primary growth rate and even increases peculiarly with $p_z$.
This secondary instability was discussed 
as nonlinear effects of the primary instability in terms of the re-summed 
self-energy diagrams in Ref.~\cite{BergesSS2007}. In fact, 
this simulation is quite relevant for physics of the glasma
in heavy-ion collisions. The simulation was performed in the 
Cartesian coordinates without expansion (unlike the glasma), 
but adopted an extremely anisotropic initial condition in 
the momentum space which was strongly inspired by 
the CGC initial condition. More explicitly, the initial 
configuration was stochastically generated according to the 
distribution (in the temporal gauge)
\begin{align}
\left\langle \left\vert A_j^a(t=0,{\boldsymbol p})
\right\vert^2\right\rangle
=
\frac{C}{(2\pi)^{3/2}\varDelta^2 \varDelta_z}
\exp\left \{
-\frac{p_z^2}{2 \varDelta_z^2}
-\frac{{\boldsymbol p}_\perp^2}{2 \varDelta^2}
\right \}\, ,
\label{eq:dist}
\end{align}
with the color index $a=1, 2, 3$ and the Lorentz index 
$j=x, y, z$. 
The normalization parameter 
$C$ is fixed by the energy density. 
The variance $\varDelta^2$ for the transverse momentum 
$p_\perp=\sqrt{p_x^2+p_y^2}$ may be identified with 
the saturation scale $Q_s^2$, while $\varDelta_z$ 
for the longitudinal momentum $p_z$ is taken as small enough 
$\varDelta_z \ll \varDelta$ to give a $\delta(p_z)$-like 
distribution practically.
We emphasize here that {\it the initial configuration thus 
generated will have a nontrivial correlation over the 
transverse distance $\sim 1/Q_s$ while it will be very smooth 
in the $z$ direction}, which is quite similar to the longitudinal 
flux tubes with rapidity-dependent soft fluctuations in the 
glasma \cite{RomatschkeV2005,FujiiI2008}. 
Therefore, what was done in Ref.~\cite{BergesSS2007}
essentially corresponds to the numerical simulation of a
{\it non-expanding glasma}.

However, there are two points on which we have to be careful in 
comparing this numerical simulation with the expanding glasma. 
First, the actual simulation starts with the supplementary condition
$E^a_j=-\dot A^a_j=0$ consistently with the Gauss law. 
This means that the initial configuration is 
{\it purely color magnetic} in contrast with the expanding 
glasma which has initially both color electric and magnetic 
flux tubes. Second, although the magnetic fields generated by the 
distribution (\ref{eq:dist}) are almost homogeneous in the 
longitudinal direction, there is no preferred direction 
of the fields in general. This is in contrast with the 
expanding glasma whose dominant components are longitudinal.
Still, it is certainly possible for some of the flux tubes 
to have longitudinal polarization in the non-expanding glasma.

In the present paper, we shall discuss a possibility 
 that the primary and secondary instabilities found in 
Ref.~\cite{BergesSS2007} are both induced by the N-O type 
instability studied in \cite{Iwazaki2008,FujiiI2008}. 
The `primary' instability occurs in a single flux tube 
in which there is a strong magnetic field in the 
longitudinal direction. This analysis is essentially 
equivalent to the case with the expanding glasma 
\cite{Iwazaki2008,FujiiI2008}. We will further argue 
that the `primary' N-O instability will induce a color 
electric current in the $z$ direction parallel to the 
original magnetic field, which then generates an azimuthal 
magnetic field around the current owing to the Amp\`ere law. 
This new magnetic field can become strong enough to cause 
the `secondary' N-O instability. The $p_z$ dependence of 
this subsequent instability turns out to be consistent 
with the findings in Ref.~\cite{BergesSS2007}.

The present paper is organized as follows. 
In the next section, we first explain the N-O instability of 
a homogeneous magnetic field directed to the $z$ direction
as an approximation to the magnetic flux tube. 
We point out qualitative similarities with the primary 
instability found numerically in Ref.~\cite{BergesSS2007}. 
Then we argue how the color electric current is induced
along the $z$ direction by the unstable modes. 
Stability analysis of the azimuthal color 
magnetic field generated by the induced current is 
presented in Section 3. Section 4 is devoted to summary and discussions.

\section{The N-O instability and induced current}

In this section, we first review the N-O instability 
of a homogeneous magnetic field in the SU(2) Yang-Mills theory 
which corresponds to a simplified situation of the magnetic 
flux tube in the non-expanding glasma. For technical reasons, 
we formulate everything in the Abelian decomposed representation 
as in the original presentation by Nielsen and Olesen~\cite{nielsen}. 
We then discuss the consequences of this N-O instability.

\subsection{The N-O instability in the Abelian 
decomposed representation}

We study the SU(2) Yang-Mills theory in the temporal gauge $A_0^a=0$.
It has been well known that the configuration of a homogeneous
magnetic field in non-Abelian gauge theories exhibits 
instability with respect to fluctuations perpendicular to 
the magnetic field in the color space. This is called the 
Nielsen-Olesen (N-O) instability \cite{nielsen}. 
This can be most easily understood when we recast the theory 
in terms of the U(1) ``electromagnetic field" 
$A_\mu \equiv A_\mu^3$ and the ``charged vector fields" 
$\phi_{\mu}\equiv (A_{\mu}^1+iA_{\mu}^2)/\sqrt{2}$
\ : 
\begin{align}
{\mathcal  L}=&
-\tfrac{1}{4} F^a_{\mu\nu}F^{a\mu\nu}
\nonumber \\
=&
-\tfrac{1}{4}f_{\mu\nu}f^{\mu\nu}
-
\tfrac{1}{2}
|D_{\mu}\phi_{\nu}-D_{\nu}\phi_{\mu}|^2
+
ig f^{\mu\nu} \phi_{\mu}^{*}\phi_{\nu}
+\tfrac{1}{4}g^2(\phi_{\mu}\phi_{\nu}^{*}-
\phi_{\nu}\phi_{\mu}^{*})^2
\; ,
\label{eq:Lag}
\end{align}
where we introduced the U(1) field strength 
$f_{\mu\nu}= \partial_\mu A_\nu -\partial_\nu A_\mu$ and
the corresponding covariant derivative
$D_{\mu}=\partial_{\mu}+igA_{\mu}$.
Notice that we can always gauge transform a homogeneous 
magnetic field to the 3rd color direction so that it can be 
represented by the U(1) gauge field. Then, we treat 
the U(1) gauge field $A_\mu$ as a {\it background} field, 
while the charged vector fields $\phi_\mu$ as small 
{\it fluctuations}.\!\footnote{We ignore the fluctuation of the U(1) 
gauge field since it is stable.} 
Note that the fluctuations can in principle vary 
in the $z$ direction, while the background field does not.

We consider a uniform, and time-independent color magnetic 
field $B>0$ in the $z$ direction which can be generated by 
$A^j=(- \tfrac{1}{2}B y, \tfrac{1}{2}B x, 0)$. 
The $\phi$-dependent part of the energy density is now written as
(in the temporal gauge $A_0=\phi_0=0$)
\begin{align}
\mathcal{E}_{\phi}=&
|\dot \phi^i|^2
+\tfrac{1}{2}
|D^{i}\phi^{j}-D^{j}\phi^{i}|^2
- i g f^{ij} \phi^{i*}\phi^j
-\tfrac{1}{4}g^2(\phi^{i *}\phi^j-\phi^{i}\phi^{j*})^2
\nonumber \\
=&
  |\dot \phi^+|^2 
+ |\dot \phi^-|^2 
+ |{\boldsymbol D}\phi^{+}|^2
+ |{\boldsymbol D}\phi^{-}|^2
+ 2g B |\phi^+|^2 
- 2g B |\phi^-|^2 
\nonumber \\
&
+ |\dot \phi^z|^2 
+ |{\boldsymbol D}\phi^{z}|^2
-\tfrac{1}{4}g^2(\phi^{i *}\phi^j-\phi^i\phi^{j*})^2
\label{eq:E}
\end{align}
with ${\boldsymbol D}= {\boldsymbol \nabla} 
- ig {\boldsymbol A}$ and
$\phi^{\pm}=\tfrac{1}{\sqrt{2}}(\phi^x \pm i\phi^y)$
in the `spin' basis. In the second line we neglected 
the surface term and exploited, for simplicity,\!\footnote{In 
fact, we are able to obtain the same results without 
imposing this condition.}
the condition $D_i\phi^i=0$,  which is consistent 
with the non-Abelian part of the Gauss law of the SU(2) gauge theory 
\begin{equation}
 D_i \dot \phi^i + ig \dot A^i\phi^i=0\; ,
\label{eq:gauss_phi}
\end{equation}
since the background field is now time-independent $\dot A^i = 0$.
Furthermore, the physical solution $\phi^i$ to the Yang-Mills equation
must satisfy the Abelian part of the Gauss law 
\begin{equation}
\partial_i \dot A^i
-ig(\phi^i \dot \phi^{i *}-\phi^{i *} \dot \phi^i)
=0 
\; ,
\label{eq:gauss_A}
\end{equation}
which simplifies to
$\phi^i \dot \phi^{i *}-\phi^{i *} \dot \phi^i=0$
when $\dot A^i=0$.
We stress here that the SU(2) gauge field theory inevitably 
involves the {\it non-minimal spin coupling}  
$[D^\mu, D^\nu]\phi_{\mu}^{*}\phi_{\nu}\sim 
igf^{\mu\nu} \phi_{\mu}^{*}\phi_{\nu}$
between $A_\mu$ and $\phi_\mu$ in addition to the minimal ones 
$-\frac12 |D_\mu \phi_\nu|^2$ 
when decomposed with respect to the U(1) subgroup.
While this non-minimal coupling entails mixing of $\phi^x$ 
and $\phi^y$, it can be diagonalized 
by the introduction of $\phi^\pm$ as seen in the second 
line of Eq.~(\ref{eq:E}). Thus, in a magnetic field,
the charged vector fields $\phi ^\pm$ acquire the ``Zeeman''
energy $\pm 2gB$ through this non-minimal coupling.
This is the origin of the N-O instability as we will 
see below.

Within the linear approximation with respect to the fluctuations 
$\phi^\pm$, we find the equation of motion for $\phi^{\pm}$ 
\begin{align}
\left(\partial_t^2 - {\boldsymbol D}^2 \pm 2gB\right)\phi^\pm=0
\, ,
\label{eq:eom}
\end{align}
with ${\boldsymbol D}^2=\nabla^2 - \frac14 g^2 B^2(x^2+y^2) 
+ig B (y\partial_x-x\partial_y)$. We can solve it in a similar 
way as in the quantum Hall effects and obtain the 
eigenfrequencies with the longitudinal momentum $p_z$ as
\begin{align}
\omega^2 = p_z^2 + gB ( 2n + |m| - m + 1) \pm 2gB\, ,
\label{eq:eigenfreq}
\end{align} 
where $n=0,1,2,\cdots$ is the principal quantum number of a 
two-dimensional harmonic oscillator and $m=0, \pm 1, \pm 2, \cdots$ 
is the eigenvalue of two-dimensional angular momentum 
$L_z=i(x\del_y -y\del_x)$. 
The factor in the brackets can be written as 
$2n+|m|-m+1 \equiv 2N+1$ 
with $N=0,1,2,\cdots$ specifying the Landau levels. Each Landau level is 
degenerate in $m \ge 0$, but we simply set 
$m=0$ since it is irrelevant to the discussion below. 
Note that the frequency for the lowest 
Landau level $N=n=0$ of $\phi^-$ (and $\phi^{-*}\equiv 
(\phi^{x*}+ i\phi^{y*})/\sqrt2$, too) allows pure imaginary 
values for small $p_z$ satisfying $p_z^2 < gB$ due to the 
non-minimal coupling, which indicates that the fluctuation 
grows exponentially $\phi^- \propto {\rm e}^{\gamma t}$ 
with the growth rate $\gamma = |{\rm Im}\ \omega|$. 
This is the instability Nielsen and Olesen pointed out some 
time ago \cite{nielsen}. 
In this ideal case of the uniform 
magnetic field, $\gamma$ is given by 
\begin{equation}
\gamma(p_z) = |{\rm Im}\ \omega| = \sqrt{gB-p_z^2}\, .
\label{eq:growth_rate}
\end{equation}
Clearly, the fastest instability occurs at $p_z=0$ 
with the maximum growth rate $\gamma_{\rm max}=\sqrt{gB}$. 
Here, we should notice 
that existence of the unstable mode at $p_z=0$ is 
a unique feature of the N-O instability in clear 
contrast with the Weibel instability, which is familiar 
in the plasma-instability scenario. 
In fact, the Weibel instability takes place when 
charged particles with an anisotropic distribution move 
in a soft {\it inhomogeneous} magnetic field with
the `minimal' particle-field coupling \cite{Plasma_instability}. 
Inhomogeneity of the soft magnetic field is indispensable 
for an anisotropic distribution of the charged particles
to show filamentation along this very inhomogeneity
inducing a current, which then creates a magnetic field additively 
to the original field. 
Thus, the Weibel instability cannot occur in the 
homogeneous limit $p_z\to 0$.

The simplest situation discussed above still has some relevance 
to the inhomogeneous background magnetic field on the 
transverse ($xy$) plane as far as the typical modulating 
distance $R_\perp$ and the typical strength $B$ of the magnetic 
field satisfy the condition $R_\perp\simge 1/\sqrt{gB}$. 
This is because the unstable mode in a 
uniform magnetic field $B$ is localized in a transverse region of 
a size $1/\sqrt{gB}$ as is evident from the explicit form of the 
solution\footnote{This solution satisfies the Abelian part of the 
Gauss law (\ref{eq:gauss_A}).} (in the $p_z$ representation):
\begin{equation}
\widetilde \phi^-_{\rm unstable}(t,x_\perp,p_z) 
\propto {\rm exp}\left\{\gamma(p_z) t 
-\frac{1}{4}gB x_\perp^2 \right\} \, .
\label{eq:solution}
\end{equation}
As discussed in Introduction, the expanding 
glasma in heavy-ion collisions is initially an assembly of many 
longitudinal color flux tubes whose transverse correlation 
length is typically given by $Q_s^{-1}$. Since the color 
magnetic field inside the flux tube is roughly estimated 
as $B\sim {\mathcal O}(Q_s^2/g)$, 
the transverse size of the unstable mode of
the N-O instability is also 
given by $1/\sqrt{gB}\sim Q_s^{-1}$. Therefore, in this 
case, the condition $R_\perp\simge 1/\sqrt{gB}$ is 
merginally satisfied, indicating the relevance of the 
N-O instability in the expanding glasma \cite{FujiiI2008}. 
The same should be true for the glasma without expansion 
if the similar estimate for the color magnetic field strength holds.
In fact, as mentioned in Introduction, 
the numerical simulation in Ref.~\cite{BergesSS2007} was 
performed choosing the initial condition similar to that 
of the expanding glasma. Therefore, it seems plausible that the 
primary instability found in Ref.~\cite{BergesSS2007} 
is due to the Nielsen-Olesen instability. The 
observed growth rate of the primary instability 
indeed shows very similar $p_z$ dependence as 
Eq.~(\ref{eq:growth_rate}), especially 
it remains {\it nonzero} at $p_z=0$.

A more rigorous analysis would require to solve Eq.~(\ref{eq:eom}) 
numerically with the magnetic field $B$ 
generated randomly on the transverse plane via Eq.~(\ref{eq:dist}). 
Then a negative eigenvalue $\omega^2$ of a solution
indicates an unstable mode. Although existence of an unstable 
mode in such a generic background is not obvious at all,
it is quite reasonable that the (largest) growth rate of 
the unstable mode, if exists, will be smaller than the 
estimate $\sqrt{g|B_{\rm max}|}$ where $|B_{\rm max}|$ is 
the maximum value of the inhomogeneous magnetic field.
This can be intuitively understood if one recognizes that 
the magnetic field $B$ corresponds to a `potential' in 
Eq.~(\ref{eq:eom}): a positive (negative) $B$ works as an 
attractive (repulsive) potential well for the fluctuation 
field $\phi^-$ and the unstable modes are regarded as 
`bound states' trapped in this potential. If one considers 
a bound state (for small $p_z$) in a single potential well 
with a certain depth, the `binding energy' becomes smaller 
in general as the potential gets more localized.
In other words, if we define an effective magnetic field 
$B_{\rm eff}$ with the (largest) negative eigenvalue of 
$\omega^2$ as $\omega^2 \equiv -g B_{\rm eff}$, 
then we have
$B_{\rm eff} < |B_{\rm max}|$. 
Furthermore, we may expect that the qualitative dependence of 
the growth rate on the momentum $p_z$ will be unchanged 
from the uniform field case. This is because
the momentum $p_z$ always enters the 
eigenfrequency $\omega^2$ in the same way as in 
Eq.~(\ref{eq:eigenfreq}), 
as far as there is no $p_z$ dependence in the background magnetic field.
After all, the $p_z$ dependence 
of the growth rate $\gamma$ in an inhomogeneous 
magnetic field will be represented\cite{Iwazaki2008} with 
the effective magnetic field as 
\begin{equation}
\gamma(p_z) \simeq \sqrt{gB_{\rm eff}-p_z^2}\, ,
\end{equation}
and the maximum momentum for the instability will 
be given by $p_z^{\rm max}=\sqrt{gB_{\rm eff}}$.

\subsection{Induced current and the azimuthal magnetic field}

The exponential growth of the unstable modes cannot last 
forever. When the amplitude of the unstable fluctuation $\phi^-$ becomes
sizeable, we need to include the effects of nonlinear interaction 
between fluctuations which were ignored in the stability analysis 
of the homogeneous magnetic field. In fact, the charged 
matter fields have a double-well potential similar to the 
Higgs field (see Eq.~(\ref{eq:E})), and thus it is quite 
reasonable that the instability will cease when the fluctuation 
becomes as large as $\phi^-\sim \sqrt{8B/g}$, corresponding to the 
bottom of the potential. However, this is not the end of the story. 
In this subsection, we discuss the consequences of 
this enhanced fluctuations.

We claim that the enhanced fluctuations will induce 
a large color electric current in the longitudinal direction, 
which then generates an azimuthal magnetic field according to the 
Amp\`ere law.\footnote{Of course, the $x$ and $y$ components 
of the current are also enhanced by the N-O instability, 
but they generate a magnetic field which helps to cancel 
the original magnetic field in the $z$ direction.}
Here we define the `induced U(1) current' 
for the U(1) gauge field in the Lagrangian: 
$J^\mu = \delta {\mathcal L}/\delta A_\mu$. 
Its $z$-component is given by 
\begin{equation}
J^z=\sum_{\alpha=\pm}
ig\left\{ 
\phi^{\alpha *} D^z \phi^{\alpha} 
- (D^z\phi^{\alpha})^{\ast}\phi^{\alpha}
\right\}\, .
\label{eq:current}
\end{equation}
Being made of the {\it charged} fluctuations 
$\phi^\pm$, this current flows in the presence of a 
color electric field in the $z$ direction. This also 
implies that the current will be enhanced if the 
fluctuations grow exponentially. 
Recall that the unstable mode can have longitudinal 
momentum $p^z \simle \sqrt{gB}$, and thus one may count 
the order of the covariant derivative $D^z$ in the current 
as $D^z\sim {\cal O}(\sqrt{gB})$. If the 
fluctuation grows up to the same order as its saturation value 
$\phi^-\sim {\cal O}(\sqrt{B/g})$, magnitude of the 
induced current can be roughly evaluated as 
\BQ
J^z \sim {\mathcal O}(g \cdot \sqrt{gB}\cdot {B}/{g}) 
= {\mathcal O}((gB)^{3/2}/g)\sim {\mathcal O}(Q_s^3/g)\, ,
\label{eq:current_order}
\EQ
where we have used an estimate $\sqrt{gB}\sim Q_s$. 
Note that this current can be parametrically very large. 

The induced current in the $z$ direction 
creates azimuthal magnetic field $B^\theta$ around it according to the 
Amp\`ere law: ${\boldsymbol J}={\rm rot}\, {\boldsymbol B}$ 
(in the present case $J^z=\frac{1}{r}\del_r(rB^\theta)$ 
if one considers only the $z$ component), where  
 the azimuthal magnetic field $B^\theta$ 
can be as strong as the initial magnetic field: 
\BQ
B^\theta = {\mathcal O} (Q_s^2/g)\, .
\label{eq:Btheta}
\EQ
For a localized current $J^z>0$ in a finite 
region of the size $1/Q_s$ on the transverse plane, 
the Amp\`ere law implies that
the azimuthal magnetic field becomes maximum at the brim of
the current with the strength estimated as in 
 Eq.~(\ref{eq:Btheta}), and it decays only slowly as 
$B^\theta (r) \propto 1/r$ outside the region of the nonzero 
current. 
In the next section, we argue that thus generated 
azimuthal magnetic field  
is again unstable against the Nielsen-Olesen mechanism.

Admittedly, our discussion on the induced current 
(\ref{eq:current_order}) 
presented here is only order-of-magnitude estimate. 
More accurate evaluation of the magnitude 
will require an analysis of nontrivial time evolution 
of the background field (in particular, the 
longitudinal color electric field $E^z=-\del_t A^z$) due to 
coupling between the background field and the enhanced 
fluctuation, which certainly 
goes beyond the linear analysis. Also more inputs from 
non-perturbative numerical simulations will be very useful. 
One may be able to argue some possible mechanisms
to generate the induced current, but at present it is 
quite difficult to provide a solid picture for that. 
We thus leave the definite analysis as an open problem for 
future study.

\section{Effects of azimuthal magnetic field $B^\theta$}

We have seen in the previous section that the homogeneous color 
magnetic field directed to the 3rd color and to the $z$ spatial 
direction undergoes the N-O instability, and that 
the enhanced fluctuations induce a current in the $z$ direction, 
which in turn generates an azimuthal magnetic field $B^\theta$. 
In this section, we shall investigate the consequences of 
the presence of this newly created magnetic field. 
As already discussed, the induced current starts to flow 
according to the generation of longitudinal color electric 
field $E^z$. Thus, the azimuthal magnetic field will become 
stronger with increasing time. In this section, however, we treat 
$B^\theta$ as time-independent since we are interested in 
instabilities which will take place in a very short period. 
Still, it would be very helpful to investigate two different 
situations: 
(i) when $B^\theta$ is weak enough and can be treated as 
perturbation to the original magnetic field $B^z$, and 
(ii) when there is a strong $B^\theta$. In the latter case,
we will ignore the effects of $B^z$ which will be 
screened by the `primary' N-O instability. 

Stability analysis in the presence of the azimuthal magnetic field
is conveniently formulated in the cylindrical coordinates 
$(r,\theta,z)$ with $x=r\cos \theta$ and $y=r\sin \theta$. 
For simplicity, we consider a constant color magnetic field  
$\mbox{\boldmath $B$}=(B^r=0,B^{\theta},B^z)$  
which is generated\footnote{
$B^r=\frac{1}{r}\del_\theta A^z - \del_z A^\theta$, 
$B^\theta=\partial_z A^r - \partial_r A^z$ and 
$B^z=\frac{1}{r}\partial_r(rA^\theta)-\frac{1}{r}\partial_\theta A^r$.}
by the gauge field : 
$(A^r, A^{\theta}, A^z)=(0,\tfrac{1}{2}rB^z, -rB^{\theta})$.
We again treat the charged vector field $\phi_\mu$ as fluctuation
and perform the linear approximation in the equations of motion for 
$\phi^i$. Then one finds three coupled equations among $\phi^r$, 
$\phi^\theta$ and $\phi^z$ (in the temporal gauge $\phi^0=0$):
\begin{align}
\label{eq:r}
&
\partial_t^2 \phi^r
-{\boldsymbol D}^2\phi^r+\frac{2}{r}D_\theta \phi^\theta
+\frac{1}{r^2}\phi^{r}
+2igB^z\phi^{\theta}
-2igB^{\theta}\phi^z=0\, ,	 \\
\label{eq:theta}
&
\partial_t^2 \phi^{\theta}
-{\boldsymbol D}^2\phi^{\theta}-\frac{2}{r} D_\theta \phi^r
+\frac{1}{r^2}\phi^{\theta}
-2igB^z\phi^r=0\, , \\
\label{eq:z}
&
\partial_t^2 \phi^z
-{\boldsymbol D}^2\phi^z+2igB^{\theta}\phi^r=0\, ,
\end{align} 
where ${\boldsymbol D}^2=D_z^2 +\partial_r^2 
+ \frac{1}{r}\partial_r +D_\theta^2$ 
and $D_\theta=\frac{1}{r}\partial_\theta -igA^\theta$.
Note that it is the azimuthal field $B^\theta$ which couples 
the field $\phi^z$ to the transverse field $\phi^r$.
In fact, when the azimuthal magnetic field is absent 
$B^\theta=0$, Eqs.~(\ref{eq:r}) and (\ref{eq:theta}) reduce 
to Eq.~(\ref{eq:eom}) for 
$\phi^\pm =(\phi^r \pm i \phi^\theta)e^{\pm i \theta}/\sqrt{2}$.

\subsection{$B^\theta$ as perturbation to $B^z$}

Before we present the stability analysis of the azimuthal 
magnetic field, let us examine how the results obtained in 
the previous section will change in the presence of small 
perturbation of $B^\theta$. First of all, 
we rewrite Eqs.~(\ref{eq:r}) and (\ref{eq:theta}) 
in terms of $\widetilde \phi^\pm$ and $\widetilde \phi^z$ (in the 
$p_z$ representation) as 
\begin{align}
\left \{
\partial_t^2 +\left({p_z} - gB^\theta r\right)^2 
-\frac{1}{r}\partial_r (r \partial_r)
-D_\theta^2 \pm 2g B^z 
\right \}\widetilde \phi^{\pm}
=& 2igB^\theta \widetilde \phi^{z}\, .
\end{align}
Compared with Eq.~(\ref{eq:eom}), one indeed verifies 
that the azimuthal field $B^\theta$ adds the coupling with 
$\phi^z$ and shifts the longitudinal momentum $p_z$ 
in our gauge. Consider the unstable fluctuation $\widetilde \phi^-$
and ignore the stable one $\widetilde \phi^z$. 
When the azimuthal magnetic field is much weaker than 
the longitudinal magnetic field $B^\theta \ll B^z$, 
we take only the term linearly dependent on $B^\theta$: 
\BQ
\left\{\left(\del_t^2 - 
{\boldsymbol D}_{(0)}^2 -2gB^z \right) 
- 2gB^\theta r p_z \right\}\widetilde \phi^-=0\, ,
\EQ
where we have introduced 
${\boldsymbol D}_{(0)}^2=\nabla_\perp^2-p_z^2 
-\frac{1}{4}g^2 (B^z)^2 r^2$ for zero angular momentum states,
with the subscript (0) indicating that $B^\theta=0$.
Taking the last term in the curly brackets as perturbation, 
we find that the lowest
eigenfrequency (squared) is modified in the leading order of 
perturbation as  
\BQ
\omega^2_{n=0} = p_z^2 -gB^z -2gB^\theta \langle r\rangle\, p_z\, ,
\EQ
where the average $\langle r \rangle$ is taken over
the unperturbed solution given by Eq.~(\ref{eq:solution}), i.e., 
$\langle r \rangle = \int_0^\infty r dr  
\widetilde \phi^{-*} r \widetilde \phi^-/\int_0^\infty r dr  
\widetilde \phi^{-*} \widetilde \phi^- \propto 1/\sqrt{gB^z}$.
This eigenvalue $\omega^2_{n=0}$
can become negative for the momentum 
$p_z$ such that
\BQ
-\sqrt{gB^z} + gB^\theta \langle r\rangle <p_z < 
\sqrt{gB^z} + gB^\theta \langle r\rangle \, ,
\EQ
yielding instability with the growth rate  
\BQ
\gamma(p_z)\simeq \sqrt{gB^z+2gB^\theta 
\langle r \rangle p_z -p_z^2}\, ,
\EQ
where we have ignored the contribution quadratic in $B^\theta$. 
The maximum momentum for the instability is given by 
$p_z^{\rm max}\simeq \sqrt{gB^z} + gB^\theta \langle r\rangle
= \sqrt{gB^z} (1+ c B^\theta/B^z)$ with $c=\sqrt{\pi/2}$,
and the unstable region in the $p_z$ space slightly 
shifts to larger $p_z$. On the other hand, the maximum value of 
the growth rate $\gamma=|{\rm Im}\, \omega |$ is still given by 
$\sqrt{gB^z}$ in the leading order perturbation. A schematic 
picture of the growth rate as a function of $p_z$ is shown in 
Fig.~1.

Let us summarize here our understanding of the `primary' N-O 
instability. The simplest picture with a constant 
magnetic field $B^z$ should be modified at two points: 
(i) inhomogeneity of the magnetic field on the transverse 
plane, and (ii) the presence of $B^\theta$. The first point
works to diminish the growth rate, and the second one 
works to shift the instability region in the $p_z$ space 
slightly to the right, to form an oval shape. These two 
effects seem to be qualitatively in agreement with the 
primary instability obtained in the numerical simulation 
\cite{BergesSS2007}.

\begin{figure}[t]
\begin{center}
\includegraphics[width=10.5cm]{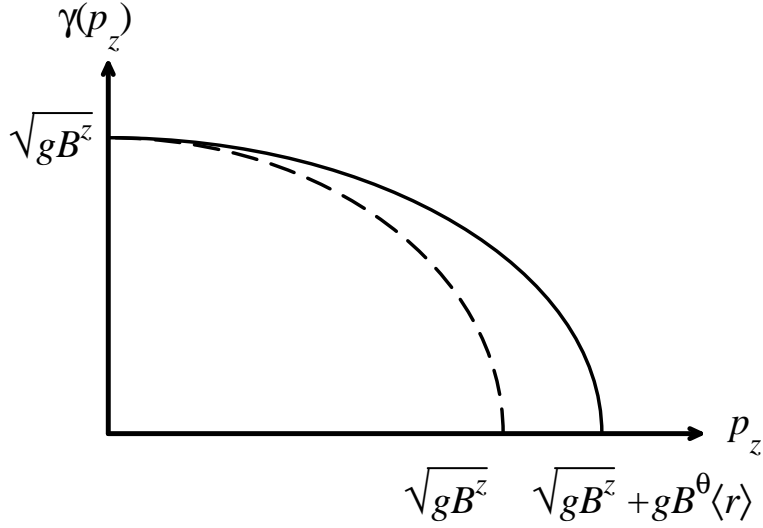}
\caption{A schematic picture of the 
growth rate $\gamma= |\text{Im}\, \omega|$ 
of the `primary' N-O instability. Solid (dashed) line is the 
growth rate with (without) a weak azimuthal magnetic field.}
\vspace{5mm}
\label{fig:primary}
\end{center}
\end{figure}

\subsection{Instability of the azimuthal magnetic field 
$B^\theta$}

\begin{figure}[t]
\begin{center}
\includegraphics[width=8.5cm,angle=-90]{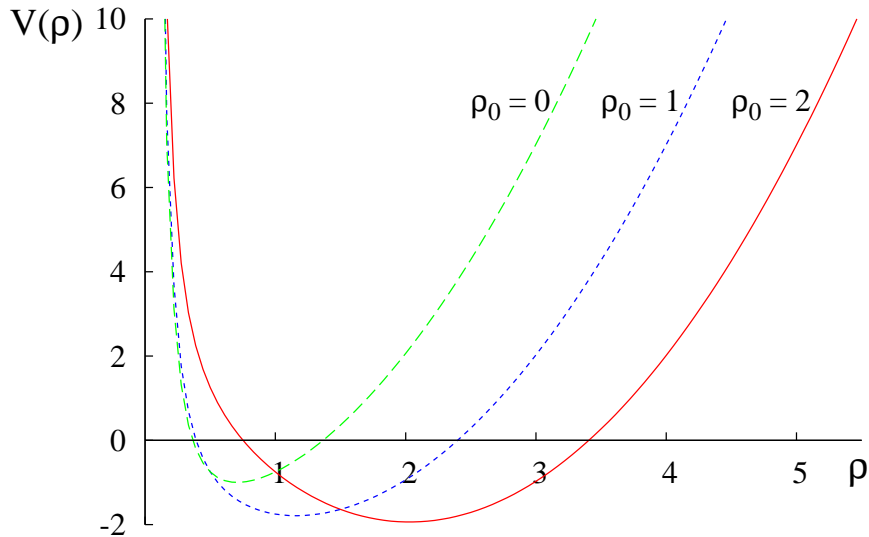}
\vspace{-5mm}
\caption{Potential $V(\rho)$ for three different values of 
$\rho_0$. The potential minimum moves outwards with increasing 
$\rho_0$.}\vspace{5mm}
\label{fig:potential}
\end{center}
\end{figure}

Let us finally present the stability analysis of 
the azimuthal magnetic field $B^\theta$. In order to 
understand the effects of $B^\theta$, we consider 
the simplest case with $B^z=0$ which is however 
reasonable because 
the original magnetic field will be weakened by the enhanced 
fluctuations. Then, Eqs.~(\ref{eq:r}) and (\ref{eq:z}) are 
nearly diagonalized
by combining the fields as
$\varphi^{\pm}=( \phi^z \pm i \phi^r )/\sqrt{2}$ :
\begin{align}
\label{eq:varphi+}
&
\partial_t^2 \varphi^+ 
 -{\boldsymbol D}^2\varphi^+ 
+ \frac{1}{2r^2}\left(\varphi^+ - \varphi^-\right)
+2gB^{\theta}\varphi^+ = 0\, ,\\
\label{eq:varphi-}
&
\partial_t^2\varphi^- 
 -{\boldsymbol D}^2\varphi^- 
- \frac{1}{2r^2}\left(\varphi^+ - \varphi^-\right)
-2gB^{\theta}\varphi^- = 0
\, ,
\end{align}
where we have again restricted ourselves to the 
states with $\partial_\theta =0$ and the operator 
${\boldsymbol D}^2$
is now given by ${\boldsymbol D}^2=D_z^2+\del_r^2 + \frac{1}{r}\del_r$.
When $B^\theta>0$, the structure of these equations suggests that 
instabilities will occur for $\varphi^-$ due to the negative 
potential term $-2gB^\theta$. Therefore, we solve the equation 
for $\varphi^-$ assuming that $\varphi^+$ is small compared to 
$\varphi^-$. By setting $\varphi^+=0$, one finds an equation 
for $\widetilde \varphi^- \propto e^{-i\omega t + ip_z z}$ :
\begin{align}
\Bigl\{-\frac{1}{r}\frac{d}{d r} r \frac{d}{d r}
+ \left(p_z - gB^{\theta}r\right)^2  
+ \frac{1}{2r^2}-2gB^{\theta}\Bigr\}\, \widetilde \varphi^-(r) =
\omega^2\,  \widetilde \varphi^-(r)
\, .
\label{eq:varphi_app}
\end{align}
This is the equation to be solved. Note that the equation has 
only one parameter $\rho_0\equiv p_z/\sqrt{gB^\theta}$, 
which becomes evident if one rewrites it 
in terms of a dimensionless variable $\rho=\sqrt{gB^\theta} r$:
\begin{equation}
\Bigl\{-\frac{1}{\rho}\frac{d}{d \rho} \rho \frac{d}{d \rho}
+ \left(\rho -\rho_0\right)^2 + \frac{1}{2\rho^2}
-2\Bigr\}\, \widetilde \varphi^- =
\epsilon\,  \widetilde \varphi^-\, ,
\label{eq:reduced2d}
\end{equation}
where $\epsilon = \omega^2 /gB^\theta$ is the normalized 
eigenfrequency. If one further rescales the fluctuation field 
as $\widetilde \varphi^-(\rho)\equiv \, \rho^{-1/2}\, \psi (\rho)$, 
then one obtains a simple one dimensional Schr\"odinger equation:
\begin{equation}
\left(-\frac{d^2}{d\rho^2} + V(\rho)\right) \psi (\rho)=
\epsilon\, \psi(\rho)\, , 
\quad 
V(\rho)\equiv \frac{1}{4\rho^2}+(\rho-\rho_0)^2 -2\, .
\label{eq:reduced}
\end{equation}
The explicit form of the potential $V(\rho)$ is shown in 
Fig.~\ref{fig:potential} for three different values of $\rho_0$. 
The potential is singular at $\rho=0$ because of the `centrifugal' 
potential acting on the $m=0$ state, inherent to the vector field,
and takes a negative value at the minimum which moves 
outwards with increasing $\rho_0$.

Let us first consider the case with large $\rho_0$ (high $p_z$).  
In this case, the centrifugal potential is negligible around 
the minimum, and thus the equation reduces to that of a 
harmonic oscillator located at $\rho=\rho_0$. The eigenvalue 
of the harmonic oscillator then gives approximate solutions to 
the equation for large $\rho_0$, namely, $\epsilon_n=(2n+1)-2\, $ 
with $n=0, 1, 2, \cdots$. Clearly, the ground state has the 
{\it negative} eigenvalue, $\epsilon_{n=0}=-1$ because of the 
last term ``$-2$'' whose origin can be traced back to the 
non-minimal coupling between the magnetic field and the 
fluctuation in Eq.~(\ref{eq:varphi_app}). The negative eigenvalue 
$\omega^2 = -gB^\theta$ implies the existence of an unstable mode 
with the growth rate $\gamma_\theta^\infty 
=\sqrt{gB^\theta}$. Notice that this 
result has no $p_z$ dependence. This is because the longitudinal 
momentum $p_z$ (or $\rho_0$) only determines the position of 
the harmonic oscillator, but does not enter the spectrum. The 
first correction to this harmonic oscillator approximation 
comes from the centrifugal potential, and yields the following 
$p_z$ dependent growth rate ($\rho_0=p_z/\sqrt{gB^\theta}$)
\begin{equation}
\gamma_\theta (p_z) \simeq \sqrt{gB^\theta} 
\sqrt{1-\frac{1}{4\rho_0^2}}\, .
\label{eq:growth_app}
\end{equation}
Therefore, it is an increasing function of $p_z$, and  
approaches the harmonic oscillator result 
$\gamma_\theta^\infty =\sqrt{gB^\theta}$ in the limit $p_z\to \infty$. 
It should be emphasized that 
this instability occurs at larger $p_z$, in contrast to the case 
of the N-O instability in a homogeneous $B^z$. 

Next we consider the case with small $\rho_0$ (low $p_z$). 
Notice that the growth rate (\ref{eq:growth_app}) becomes smaller
for smaller $\rho_0$. In fact, the equation (\ref{eq:reduced}) 
in the limit $\rho_0=0$ can be exactly solved by the confluent 
hypergeometric function with the eigenvalues 
$\epsilon_n = 4n + \sqrt2$ for $n=0,1,2,\cdots$\cite{Landau}.
All the eigenvalues are positive, and therefore the system is 
stable when $p_z=0$. This implies that there is a minimum 
value of $p_z$ for the instability.

\begin{figure}[t]
\begin{center}
\includegraphics[width=8.5cm,angle=-90]{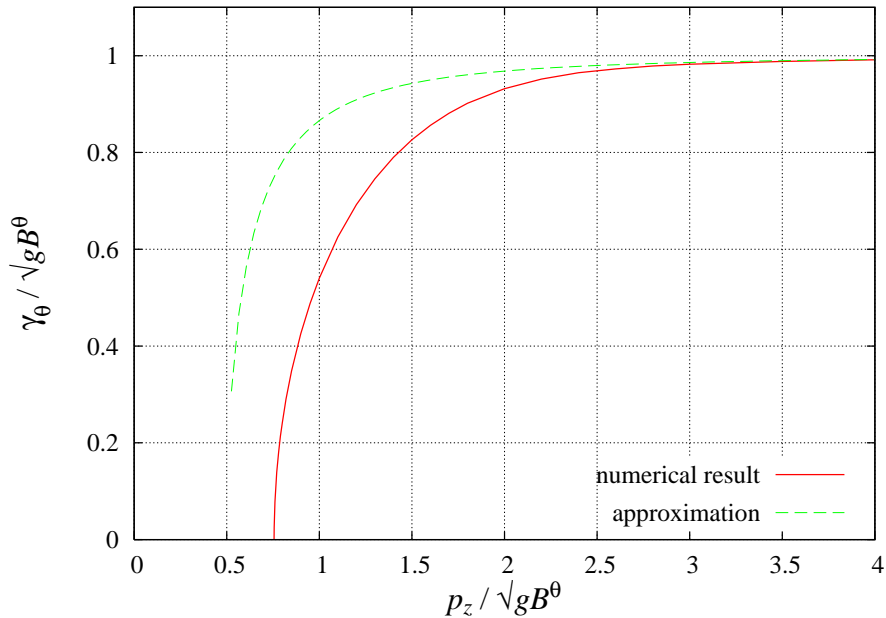}
\caption{Growth rate $\gamma_\theta (p_z)=|{\rm Im}\, \omega|$ of
the unstable mode in the azimuthal magnetic field $B^\theta$. 
Solid line: numerical result from Eq.~(\ref{eq:reduced}), 
dashed line: approximate result, Eq.~(\ref{eq:growth_app}) 
valid at large $p_z$.}
\label{fig:growthrate}\vspace{5mm}
\end{center}
\end{figure}

\begin{figure}[t]
\begin{center}
\includegraphics[width=8.5cm,angle=-90]{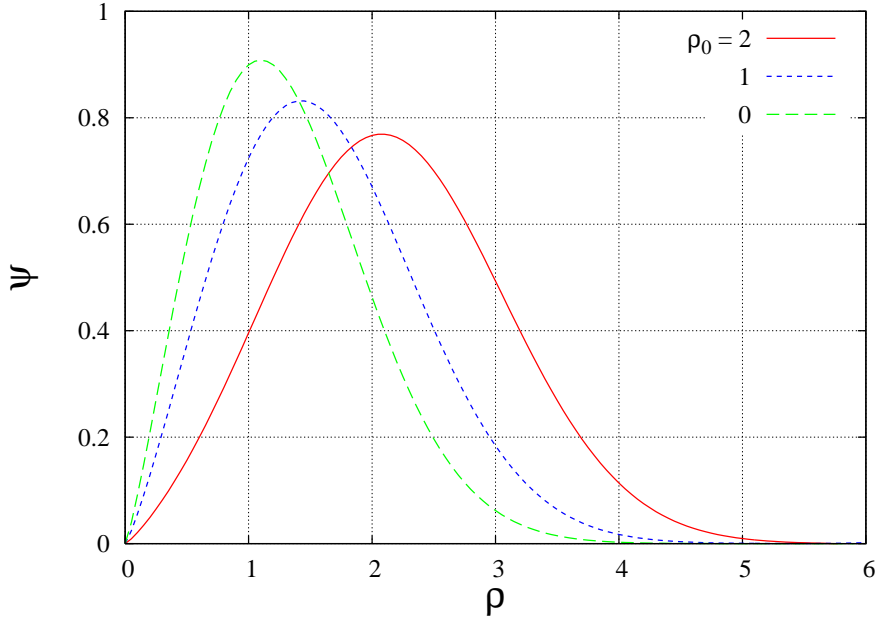}
\caption{Rescaled wavefunction $\psi(\rho)$ as a solution to 
Eq.~(\ref{eq:reduced}) for three different
potentials shown in Fig.~\ref{fig:potential}.
The case of $\rho_0=0$ is stable, while the other two 
are unstable.}
\label{fig:phi}\vspace{5mm}
\end{center}
\end{figure}

Summarizing these analytic studies, we find that the instability 
indeed occurs at relatively large $p_z$ with the growth rate 
given as an increasing function of $p_z$, while there will 
be no unstable modes for small values of $p_z$. We can directly
check that these expectations are indeed the case
by solving numerically 
Eq.~(\ref{eq:varphi_app}). The result is shown in 
Fig.~\ref{fig:growthrate} where the (normalized) growth rate 
$\gamma_\theta / \sqrt{gB^{\theta}}$ is plotted as a function of 
$p_z/\sqrt{gB^{\theta}}$. In accordance with the analytic result 
(shown as the dashed line in Fig.~\ref{fig:growthrate}), 
the instability appears only for larger $p_z/\sqrt{gB^\theta} \ge 0.755$
and the growth rate indeed increases towards the asymptotic 
value $\gamma_\theta^\infty=\sqrt{gB^\theta}$ with increasing $p_z$. 
This is the notable feature and consistent with the behavior 
observed for the secondary instability in Ref.~\cite{BergesSS2007}.
We also draw the wavefunctions $\psi(\rho)$ in Fig.~\ref{fig:phi} 
corresponding to the three different potentials shown in 
Fig.~\ref{fig:potential}. The wavefunctions are well localized 
around $\rho=\rho_0$ and vanishes at $\rho=0$. The same is true 
for the original fluctuation field 
$\widetilde \varphi^-(\rho)=\rho^{-1/2}\psi(\rho)$. 
Therefore, we expect that, as far as $p_z$ is not too large, 
our analysis with the homogeneous azimuthal magnetic field 
is not a bad approximation to more realistic cases with 
$B^\theta(\rho=0)=0$ and $B^\theta(\rho\gg 1)\sim 1/\rho$.

\vspace{5mm}

\section{Summary and discussions}

We have studied instabilities in the configuration with the
azimuthal magnetic field in the SU(2) Yang-Mills theory,
as a possible mechanism for the secondary instability observed in
the classical stochastic simulation \cite{BergesSS2007}.

The instability of an expanding glasma numerically found 
in Ref.~\cite{RomatschkeV2005} was previously 
interpreted as the Nielsen-Olesen 
instability of the longitudinally uniform color 
magnetic fields produced in between two sheets of the Color 
Glass Condensates \cite{Iwazaki2008,FujiiI2008}.
Although the simulation in Ref.~\cite{BergesSS2007} was performed 
in the Cartesian coordinates, the adopted initial condition was
extremely anisotropic in the momentum space, which is inherent 
to the collision dynamics, and we consider the primary
instability observed there has the same origin as the one in 
Ref.~\cite{RomatschkeV2005}.

We have argued that the charged fields grow up due to the 
N-O instability and are accelerated in the evolving background 
gauge field, inducing a current possibly along the $z$ axis.
This current will generates the azimuthal magnetic field,
whose strength $B^\theta$ can grow exponentially in time at 
early stage. The induction of the current and its strength
is the most speculative part of our discussion beyond the 
linear analysis.

We have shown that, once the azimuthal magnetic field becomes 
strong enough, this configuration is subsequently accompanied 
by the N-O instability with respect to another type of
fluctuation. We have analyzed this new instability in some detail,
and shown that the growth rate increases with the momentum $p_z$. 
This is in contrast with the growth rate of the primary N-O 
instability which decreases with increasing $p_z$ and vanishes 
at some value of $p_z$. The effects of the azimuthal magnetic 
fields on the primary instability are to enlarge the value of 
the momentum at which the growth rates vanish. These findings 
coincide quite well with the results in the simulation
in Ref.~\cite{BergesSS2007}.

The N-O instability scenario for the secondary instability is
generic to the Yang-Mills systems with an anisotropic configuration,
and therefore it should, in principle, apply to the case of
expanding glasmas, too. In the numerical simulation in 
Ref.~\cite{RomatschkeV2005}, however, the secondary
instability was not clearly observed. This might be because
they started the simulation with extremely small fluctuations.
On the other hand, in Ref.~\cite{BergesSS2007}, the initial
amplitudes of the fluctuations were not assumed small (actually,
this scale non-separation was one of their motivations for taking
the numerical approach). Nevertheless, it is intriguing to note
that, reexamining the pressure evolution started with relatively
larger fluctuations shown in Ref.~\cite{RomatschkeV2005}, one
can barely see a subtle kink, which may hint the secondary
instability. In fact, the increase of the longitudinal pressure
at late time can be attributed to the generation of the transverse
magnetic fields, and this is consistent with our scenario.
Another indication of the secondary instability is an abrupt
increase of the maximum longitudinal momentum $\nu_{\rm max}(\tau)$
of the unstable modes. Since the slope $C$ of the linear growth
$\nu_{\rm max}(\tau)=C\tau$ at earlier time is found to be
proportional to the growth rate \cite{Iwazaki2008,FujiiI2008},
it is plausible to relate the sudden increase of
$\nu_{\rm max}(\tau)$ to the emergence of another instability
with a larger growth rate. 
Other possible scenarios are previously discussed in 
Ref.~\cite{RomatschkeV2005}.  
It is thus a pressing subject to analyze specifically 
the course of the expanding glasma evolution from the viewpoint 
of the N-O instability.

Our analysis seems to indicate that in the glasma evolution 
the sequential appearance of instabilities play a key role 
to make the system more random and turbulent towards thermalization.
Of course, our discussion is based on the simple stability analysis
and speculation on the induced current. Especially more solid 
discussions on the generation of the azimuthal magnetic field 
and the back-reaction from the well-developed unstable mode will 
require careful analyses on the system configuration and 
evolution which will be accessible in numerical simulations. 
Moreover, 
simulations in Ref.~\cite{BergesSS2007} should involve
the Weibel instability mechanism also in principle,
although they were formulated in terms of the classical fields.
Hence, coexistence of the Weibel and the N-O instabilities
in the evolving system is worthwhile to be studied.
In this context, 
more joint efforts in analytic and numerical approaches
are to be accomplished in this field.

{\bf Acknowledgements}\\
This work was completed during the workshop on 
``{\it Non-equilibrium quantum field theories 
and dynamic critical phenomena}'' at the Yukawa Institute of 
Theoretical Physics in March 2009. 
H.F.\ and K.I.\ thank useful and extensive 
discussions with the participants, in particular with 
J\"urgen Berges. 
A.I.\ is grateful to members of KEK theory group and of Komaba
nuclear theory group for kind hospitality. 
This work is 
partly supported by Grants-in-Aid (18740169 (K.I.) and 
19540273 (H.F.)) of MEXT.


\begin{thebibliography}{99}


\bibitem{HIC}
  For example, U.~W.~Heinz and P.~F.~Kolb,
  ``{\it Two RHIC puzzles: Early thermalization and the HBT problem},''
  arXiv:~hep-ph/0204061.


\bibitem{review_Itakura}
  For a brief review, see K.~Itakura,
  Prog.\ Theor.\ Phys.\ Suppl.\  {\bf 168} (2007) 295.


\bibitem{Plasma_instability}
   For a recent review, see S.~Mrowczynski,
  PoS C {\bf POD2006} (2006) 042
  [arXiv:hep-ph/0611067].

\bibitem{bottom-up}
  R.~Baier, A.~H.~Mueller, D.~Schiff and D.~T.~Son,
  Phys.\ Lett.\  B {\bf 502} (2001) 51
  [arXiv:hep-ph/0009237].

\bibitem{nielsen}
    N.~K.~Nielsen and P.~Olesen, Nucl.\ Phys.\ {\bf B144} (1978) 376;
    Phys.\ Lett.\ {\bf 79B} (1978) 304. 
    S.~J.~Chang and N.~Weiss,
  Phys.\ Rev.\  D {\bf 20} (1979) 869.

\bibitem{Iwazaki2008}
  A.~Iwazaki,
  ``{\it Decay of Color Gauge Fields in Heavy Ion Collisions and Nielsen-Olesen
  Instability},'' to appear in Prog.\ Theor.\ Phys.\ (2009),
  arXiv:0803.0188 [hep-ph].

\bibitem{FujiiI2008}
  H.~Fujii and K.~Itakura,
  Nucl.\ Phys.\  A {\bf 809} (2008) 88 
  [arXiv:0803.0410 [hep-ph]].

\bibitem{CGC}
  For reviews, see F.~Gelis, T.~Lappi and R.~Venugopalan,
  Int.\ J.\ Mod.\ Phys.\  E {\bf 16} (2007) 2595
  [arXiv:0708.0047 [hep-ph]], E.~Iancu and R.~Venugopalan,
  ``{\it The color glass condensate and high energy scattering in QCD},''
  arXiv:hep-ph/0303204, published in ``QGP3,'' edited by R.C.~Hwa 
  and X.N.~Wang, World Scientific. 



\bibitem{Glasma_basic}
    A.~Kovner, L.~McLerran and H.~Weigert, 
    Phys. Rev. D {\bf 52} (1995) 6231; 3809.

\bibitem{LappiM2006}
  T.~Lappi and L.~McLerran,
  Nucl.\ Phys.\  A {\bf 772} (2006) 200 
  [arXiv:hep-ph/0602189].

\bibitem{RomatschkeV2005}
  P.~Romatschke and R.~Venugopalan,
  Phys.\ Rev.\ Lett.\  {\bf 96} (2006) 062302
  [arXiv:hep-ph/0510121];
  Phys.\ Rev.\  D {\bf 74} (2006) 045011 
  [arXiv:hep-ph/0605045].
\bibitem{FukuGM2006}
  K.~Fukushima, F.~Gelis and L.~McLerran,
  Nucl.\ Phys.\  A {\bf 786} (2007) 107
  [arXiv:hep-ph/0610416].


\bibitem{BergesSS2007}
  J.~Berges, S.~Scheffler and D.~Sexty,
  Phys.\ Rev.\  D {\bf 77} (2008) 034504 
  [arXiv:0712.3514 [hep-ph]].


\bibitem{BergesGSS2008}
  J.~Berges, D.~Gelfand, S.~Scheffler and D.~Sexty,
  ``{\it Simulating plasma instabilities in SU(3) gauge theory},''
  [arXiv:0812.3859 [hep-ph]].

\bibitem{Landau}
  For example, L.D.~Landau, and E.M.~Lifshitz,  
  ``{\it Quantum Mechanics: Non-Relativistic Theory,}'' 
  (Butterworth-Heinemann, 1981).

\end{thebibliography}
\end{document}